\documentclass[aps,twocolumn,preprintnumbers]{revtex4}

\usepackage{graphicx}  
\usepackage{subcaption}
\usepackage{multirow}
\usepackage{tikz}
\usepackage{fancyhdr}
\usepackage[T1]{fontenc}
\usepackage{dcolumn}   
\usepackage{bm}        
\usepackage{amsfonts}  
\usepackage{amsmath}   
\usepackage{amssymb}
\usepackage{mathrsfs}
\usepackage{comment}
\usepackage{caption}
\usepackage{caption}
\usetikzlibrary{shapes.geometric, arrows}
\tikzstyle{block} = [draw, fill=white, rectangle, minimum height=3em, minimum width=6em]
\tikzstyle{arrow} = [thick,->,>=stealth]
\makeatletter

\long\def\@makefntext#1{\parindent 1em\noindent
\hb@xt@1.8em{\hss\textsuperscript{\normalfont\@thefnmark}}#1}
\makeatother

\begin{document}

\title{Precision Mass Measurements of \textsuperscript{130}Te, \textsuperscript{130}Sn, and Their Impact on Models for R-Process Nucleosynthesis} 
\author{A. Cannon$^{a}$\footnotemark[1]\footnotemark[2]}
\author{W.S. Porter$^{a}$}
\author{A.A. Valverde$^{b,c}$}
\author{D.P. Burdette$^{b}$}
\author{A.M. Houff$^{a}$}
\author{B. Liu$^{a,b}$}
\author{A. Mitra$^{a}$}
\author{G.E. Morgan$^{b,d}$\footnotemark[3]}
\author{C. Quick$^{a}$\footnotemark[4]}
\author{D. Ray$^{b,c}$\footnotemark[2]}
\author{L. Varriano$^{b,e}$\footnotemark[5]}
\author{M. Brodeur$^{a}$}
\author{J.A. Clark$^{b,c}$}
\author{G. Savard$^{b,e}$}
\author{G.J. Mathews$^{a}$}

\affiliation{$^{a}$Department of Physics and Astronomy, University of Notre Dame, Notre Dame, Indiana 46656, USA}
\affiliation{$^{b}$Physics Division, Argonne National Laboratory, Lemont, Illinois 60439, USA}
\affiliation{$^{c}$Department of Physics and Astronomy, University of Manitoba, Winnipeg, Manitoba R3T 2N2, Canada}
\affiliation{$^{d}$Department of Physics and Astronomy, Louisiana State University, Baton Rouge, Louisiana 70803, USA}
\affiliation{$^{e}$Department of Physics, University of Chicago, Chicago, Illinois 60637, USA}

\date{\today}

\begin{abstract}
The astrophysical rapid neutron capture nucleosynthesis process (r-process) remains an active area of research due to the fact that it occurs in extreme conditions and involves reactions with exotic nuclei that are difficult to study experimentally. For the first time using the Phase-Imaging Ion Cyclotron Resonance (PI-ICR) technique, we measured the mass excesses of \textsuperscript{130}Te, \textsuperscript{130}Sn, and \textsuperscript{130}Sn\textsuperscript{m} with the Canadian Penning Trap (CPT). Our results show good agreement with previous Penning trap values obtained using the Time-of-Flight Ion Cyclotron Resonance (TOF-ICR) and the Fourier Transform Ion Cyclotron Resonance (FT-ICR) techniques, while being twice as precise for \textsuperscript{130}Sn. These new mass excesses were added to a SkyNet network calculation to determine their impact on r-process abundances and to find the best astrophysical conditions to reproduce the Solar System r-process abundance pattern. Finally, by treating  lighter and heavier elements separately, we assess the relative frequency of events producing elements in a cold versus a hot r-process scenario.
\end{abstract}

\maketitle
\footnotetext[1]{Contact author: acannon@triumf.ca}
\footnotetext[2]{Current address: TRIUMF, 4004 Wesbrook Mall, Vancouver, British Columbia V6T 2A3, Canada.}
\footnotetext[3]{Current address: Cyclotron Institute, Texas A\&M University, College Station, TX 77843, USA}
\footnotetext[4]{Current address: Department of Physics and Astronomy, University of Tennessee, Knoxville, TN 37996, USA.}
\footnotetext[5]{Current address: Center for Experimental Nuclear Physics and Astrophysics, University of Washington, Seattle, Washington 98195, USA.}
\section{\label{sec:level1}Introduction}
The rapid neutron capture process (r-process) is responsible for producing about half of all isotopes heavier than iron \cite{Arnould2007}. This nucleosynthesis process proceeds from seed nuclei that undergo successive neutron captures on timescales much faster than the timescales of beta decay near stability. The complex network calculations from which r-process abundances are extracted require both astrophysics and nuclear physics inputs. However, because the nuclei along the r-process path are very neutron-rich and unstable, it is currently difficult to produce them in the laboratory. Therefore, several critical nuclear physics inputs, such as atomic masses, need to be obtained from models that require precise and accurate data as their anchor points \cite{Porter2024}. Hence, precise and accurate mass measurements of neutron-rich isotopes near the r-process path are of great importance \cite{Mumpower2015}.

Driven by recent astrophysical observations \cite{Wang2020, Watson2019, Abbott2017} and informed by studies on how nuclear inputs affect models of these events \cite{Zhu2021, Kajino2017}, many atomic mass measurement campaigns aim towards precisely determining the masses of isotopes important for r-process nucleosynthesis \cite{VanSchelt2013,Orford2018,Vilen2018,Vilen2020, Orford2022,Jaries2024,Kimura2024,Ray2024arxiv}. 

Using these measured and modeled masses, reaction networks within r-process models can be used to simulate final isotopic abundances produced during astrophysical events. One such model is SkyNet \cite{Lippuner2017}. SkyNet gets its nuclear input data from the JINA Reaclib dataset \cite{Cyburt2010} to solve for reaction rates for a given set of initial conditions. It then propagates the network forward in time and updates the abundances as reactions occur.

In this paper, we present new measurements of the neutron-rich isotopes, \textsuperscript{130}Te and \textsuperscript{130}Sn, and of the isomer, \textsuperscript{130}Sn\textsuperscript{m}. We then perform SkyNet simulations to determine which initial astrophysical conditions best reproduce the solar r-process abundance pattern \cite{Arlandini1999}. Next, we update the masses for these nuclei as well as the Q-values for reactions involving them within SkyNet before performing more simulations to determine which astrophysical scenarios are affected by the new masses. The masses for our isotopes of interest in SkyNet's base model as well as the measured masses are shown in Table \ref{tab: masses}. Finally, we provide a quantitative determination of the relative contributions of various representative scenarios to the solar r-process abundance distributions.

\section{\label{sec:level1}Experiment}

The atomic mass measurements were performed at the Argonne Tandem Linac Accelerator System (ATLAS) facility, using the CAlifornium Rare Isotope Breeder Upgrade (CARIBU) \cite{Savard2011} facility as the ion source and the Canadian Penning Trap (CPT) mass spectrometer \cite{Savard2001} for measurements. CARIBU produces neutron-rich nuclei from an approximately $0.5$ Ci spontaneously-fissioning  $^{252}$Cf source. A large volume gas catcher \cite{Savard2003} is used to thermalize fission fragments from the source that recombine to a $1^+$ or $2^+$ charge state in a high purity helium gas before ions having a particular mass to charge ratio, A/q, are filtered using a magnetic dipole mass separator \cite{Davids2008}. This continuous ion beam is then injected into a radiofrequency quadrupole (RFQ) cooler and buncher, where it is cooled through collisions with a medium-pressure, high-purity helium gas, and ejected as bunches. These ion bunches are injected into a multi-reflection time of flight (MR-TOF) mass separator that can achieve a mass resolution of $R=m/\Delta m>100000$~\cite{Hirsh2016}. A Bradbury-Nielsen gate \cite{Ys1936} is used to select the isotope of interest. The beam is then delivered to the CPT experimental setup where it is further cooled using a linear RFQ trap before injection into the CPT for the mass measurement using the Phase-Imaging Ion Cyclotron Resonance (PI-ICR) technique. Finally, the ions are ejected from the CPT onto a position-sensitive multichannel plate (PS-MCP) detector to obtain their positions and extract the phases \cite{JAGUTZKI2002244}.

The PI-ICR technique was first developed at SHIPTRAP \cite{Eliseev2013, Eliseev2014} and is able to measure atomic mass with a relative precision of $\delta \text{m}/\text{m}\leq 10^{-8}$ \cite{Dilling2018}. As implemented at the CPT, this technique measures the cyclotron frequency,
\begin{equation}
    \label{eq: pi-icr}
    \nu_c=\frac{qB}{m},
\end{equation}
through the measurements of phase differences accumulated over a specified time in the trap using the relation:

\begin{equation}
    \label{eq: pi-icr}
    \nu_c=\frac{\Delta\phi+2\pi N}{2\pi t_{\text{acc}}},
\end{equation}
where $\Delta\phi$ is the difference between a reference phase and a final phase that the ions accumulate over a period, $t_{\text{acc}}$, after undergoing $N$ complete revolutions.

The CPT is a standard Penning Trap with hyperbolic ring and endcap electrodes. It also has a correction ring and tube electrodes and is located within a 5.7~T magnetic field. The ion motion in an ideal Penning Trap consists of three eigenmotions, the axial frequency $(\nu_z$), reduced cyclotron frequency ($\nu_+$), and magnetron frequency ($\nu_-$). $\nu_z$ is the frequency of oscillation orthogonal to the radial cyclotron motion. $\nu_+$ and $\nu_-$ are the frequencies of the radial eigenmotions of the system. In an ideal Penning trap, $\nu_c=\nu_++\nu_-$, and in general, $\nu_c>\nu_+>\nu_z>\nu_-$ \cite{RevModPhys.58.233}. 

Each phase measurement applies three RF excitations for each trap cycle. First, a $\nu_-$ dipole pulse centers the ions in the trap. Second, a $\nu_+$ dipole pulse increases the radius of the corresponding motion to excite the ions to a desired orbit. The final excitation is a $\nu_c$ quadrupole pulse converting the ions’ fast $\nu_+$ motion to the slow $\nu_-$ motion before they are ejected from the trap to the detector. This happens immediately after the $\nu_+$ dipole pulse for reference phase measurements but has a delay for final phase measurements which is the $t_{\text{acc}}$ in Equation \ref{eq: pi-icr}. A detailed explanation of the CPT electronics and measurement scheme is available in \cite{Ray2025}.

\begin{figure}[htbp]
    \centering

    \begin{minipage}[b]{0.49\columnwidth}
        \centering
        \includegraphics[width=\linewidth]{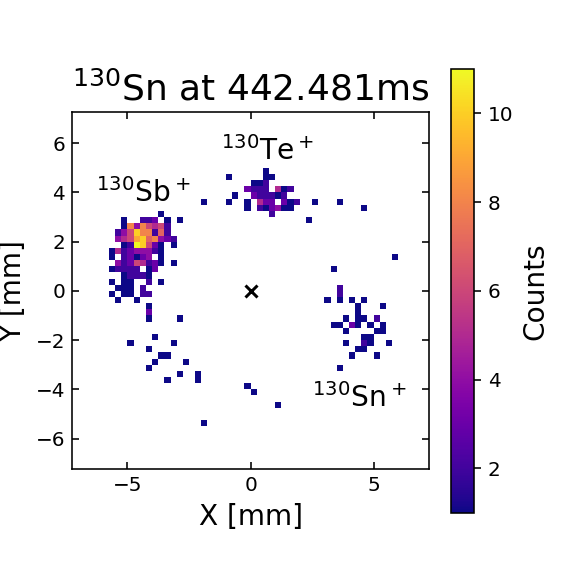}
        \centerline{(a)}
        \label{subfig:Sn130}
    \end{minipage}
    \hfill
    \begin{minipage}[b]{0.49\columnwidth}
        \centering
        \includegraphics[width=\linewidth]{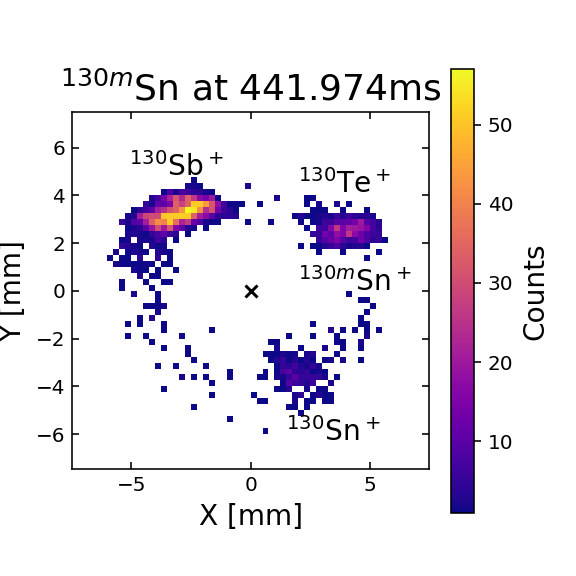}
        \centerline{(b)}
        \label{subfig:Sn130m}
    \end{minipage}

    \vspace{0.5em}

    \begin{minipage}[b]{0.49\columnwidth}
        \centering
        \includegraphics[width=\linewidth]{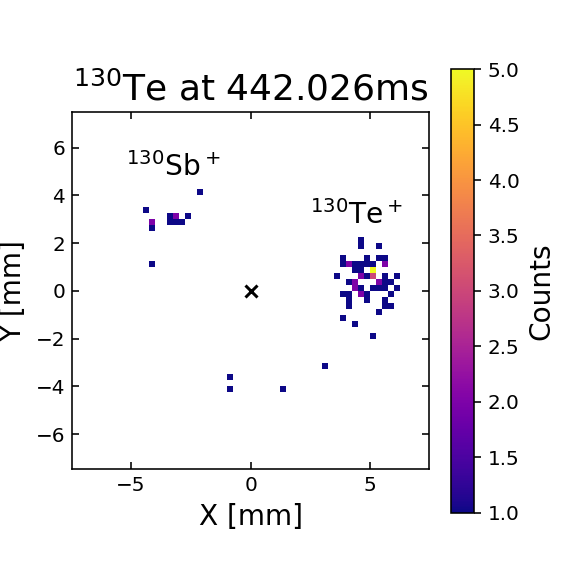}
        \centerline{(c)}
        \label{subfig:Te130}
    \end{minipage}
    \hfill
    \begin{minipage}[b]{0.49\columnwidth}
        \centering
        \includegraphics[width=\linewidth]{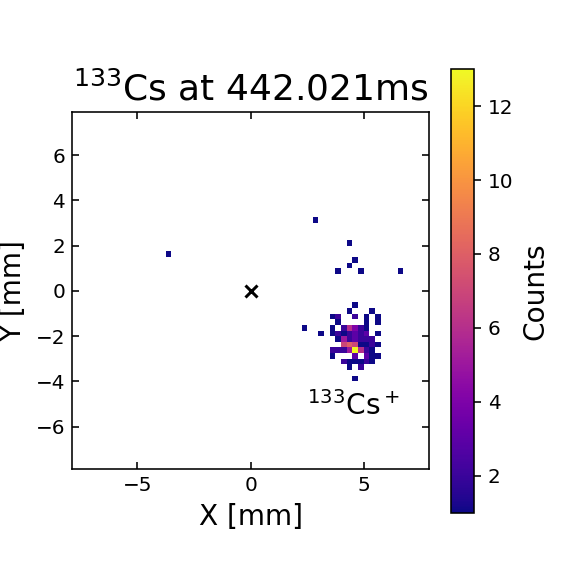}
        \centerline{(d)}
        \label{subfig:Cs133}
    \end{minipage}

    \caption{Histograms of ions detected by the PS-MCP for final phase measurements of a) \textsuperscript{130}Sn\textsuperscript{+} at $t_{\text{acc}}$=442.481ms, b) \textsuperscript{130}Sn\textsuperscript{m+} at $t_{\text{acc}}$=441.974ms, c) \textsuperscript{130}Te\textsuperscript{+} at $t_{\text{acc}}$=442.026ms, and d) \textsuperscript{133}Cs\textsuperscript{+} at $t_{\text{acc}}$=442.021ms. Contaminants for each measurement are labeled, and the trap center is indicated by an X.}
    \label{fig: Sn130}
\end{figure}
\begin{figure}[h]
    \centering
    \includegraphics[width=\linewidth]{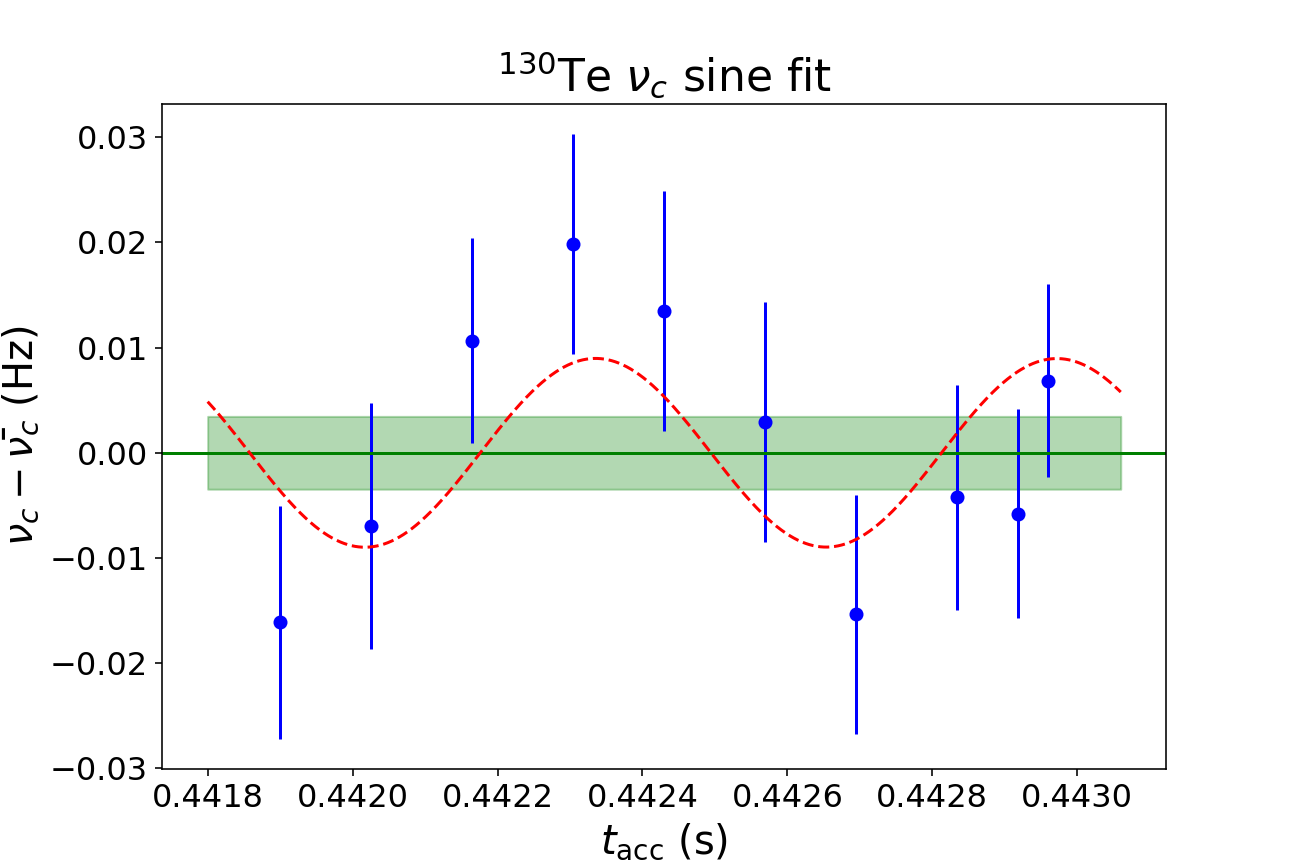}
    \caption{Measured $\nu_c$ values for $^{130}$Te from $t_{\text{acc}}=441.900$ms to $t_{\text{acc}}=442.960$ms showing the sinusoidal dependence due to the residual magnetron motion. The red dashed line represents a fit to the theoretical model described in \cite{Orford2020}, and the green horizontal line and bar represent the true $\bar{\nu_c}$ and its uncertainty.}
    \label{fig: sin}
\end{figure}

Figure \ref{fig: Sn130} shows an example of a phase measurement for each species: \textsuperscript{130}Te, \textsuperscript{130}Sn, \textsuperscript{130}Sn\textsuperscript{m}, and \textsuperscript{133}Cs. Each cluster of points on these graphs represents a different ion species. A Gaussian clustering model \cite{Weber2022} was used to find the centers of each cluster for phase measurement. This data also only represents one measurement for each species. However, due to residual $\nu_-$ motion after the initial $\nu_+$ excitation, see \cite{Ray2025}, a series of measurements at several accumulation times sampling approximately 1.5 $\nu_-$ periods must be taken. As shown in Figure \ref{fig: sin}, we can fit that data with the model relation \cite{Orford2020}:
\begin{equation}
    \nu_c(t_{\text{acc}})=\frac{p_1}{t_{\text{acc}}}\sin{(\nu_-t_{\text{acc}}+p_2)}+\bar{\nu_c}.
    \label{eq: sine model}
\end{equation}
In Equation \ref{eq: sine model}, $p_1$, $p_2$, and $\bar{\nu_c}$ are allowed to vary to fit the data, $\nu_-$ is measured beforehand, and $t_{\text{acc}}$ is the independent variable. After fitting, $\bar{\nu_c}$ is used in the final mass calculations. Since we have to take multiple measurements of the final phase for different accumulation times, we also take multiple reference phase measurements to account for any possible temporal drifts in the system. We then linearly interpolate the various reference phase measurements to find the best reference phase at the time of the final phase measurement.

\begin{table*}[t]
    \caption{Cyclotron frequency ratios and mass excesses (ME) from the present work compared with AME \cite{Wang2021}, NUBASE\cite{Kondev_2021}, and the default SkyNet masses \cite{Lippuner2017}}
    \begin{ruledtabular}
    \begin{tabular}{ccccc} 
    Species & $R=\nu_{\text{ref}}/\nu_c$ & $ME_{\text{CPT}}$ (keV) & $ME_{\text{AME}}$ (keV)& $ME_{\text{SkyNet}}$ (keV)\\
    \hline
    $^{130}$Te & $0.9774332571(171)$ & $-87354.4(2.1)$ & $-87352.960(0.011)$& -87352.9\\ 
    $^{130}$Sn & $0.9774916287(143)$ & $-80128.0(1.8)$ & $-80132.2(1.9)$& -80137.2\\
    \textsuperscript{130}Sn\textsuperscript{m} & $0.9775073638(289)$ & $-78180.0(3.6)$ & $-78185.3(1.9)$& N/A\\
    \end{tabular}
    \end{ruledtabular}
    \label{tab: masses}
\end{table*}

To extract the mass of the ion of interest, the magnetic field strength must be calibrated. To do this, we measure $\nu_c$ for a reference species of well-known mass, $m_{\text{ref}}$, giving $\nu_{\text{ref}}$. For the measurements presented here, we used $^{133}$Cs as the calibrant, and all ions were singly charged. Then, we take the ratio of $\nu_{\text{ref}}/\nu_c$, and we can calculate the atomic mass of our species of interest using
\begin{equation}
    m=\frac{\nu_{\text{ref}}(m_{\text{ref}}-m_e)}{\nu_c}+m_e,
    \label{eq: mass calc}
\end{equation}
where $m_e$ is the electron mass, and the electron binding energy is neglected.

There are several systematic effects we had to consider in our PI-ICR analysis. These are described fully in \cite{Ray2025} and \cite{Orford2020}. First, there is a correction for the presence of contaminant ions in the trap. Contaminants affect the position of the measured reference phase according to
\begin{equation}
    \phi_{\text{corr}}=2\pi t_{\text{ext}}\sum_i\chi_i(\nu_c^i-\nu_c),
    \label{eq: ratio corrention}
\end{equation}
where $t_{\text{ext}}$ is the total excitation time of the $\nu_+$ and $\nu_c$ pulses, $\chi_i$ are the population fractions of the contaminants, $\nu_c^i$ are the cyclotron frequencies of the contaminants, and $\nu_c$ is the cyclotron frequency for the ion of interest. This correction, $\phi_{\text{corr}}$, is then added to $\phi_{\text{ref}}$ before calculating the final $\Delta \phi$. This procedure is then repeated using the new $\nu_c$ until the changes in $\nu_c$ between iterations are smaller than the uncertainty.

Another source of systematic uncertainty stems from the noncircular projection of the ions' phase onto the PS-MCP that is visible in the $^{130}$Sn and $^{130}$Sn\textsuperscript{m} data in Figure \ref{fig: Sn130}. To minimize this effect, we choose accumulation times such that the spot of our ion of interest is within $\pm$10\textdegree\  of the reference ion cluster. We also apply the following correction to the calculated ratio, $\nu_{\text{ref}}/\nu_c$:
\begin{equation}
    R_{\text{C}}=R_{\text{UC}}\frac{1+b\cos{(\theta_\text{X}+\phi)}}{1+b\cos{(\theta_{\text{cal}}+\phi)}},
    \label{eq: ratio corrention}
\end{equation}
where $R_\text{C}$ and $R_{\text{UC}}$ are the corrected and uncorrected ratios, $\nu_{\text{ref}}/\nu_c$, respectively; $\theta_\text{X}$ and $\theta_{\text{cal}}$ are the average angle of the ion of interest and the calibration ion respectively; and $b$ and $\phi$ are fitted parameters found to be $3.89(32)\times10^{-8}$ and $-45.7(3.4)$\textdegree, respectively \cite{Ray2025}.

Finally, there are various systematic effects that result in a constant shift in the measured cyclotron frequency. Their combined effects result in a shift in the cyclotron frequency ratio that depends linearly on the mass difference between the species of interest and the calibrant. The magnitude of this effect is estimated to be $4.1(17)\times10^{-10}({A}/{q}-{A_{\text{cal}}}/{q_{\text{cal}}})$. This was conservatively added to the systematic uncertainty.

\begin{figure}[h]
    \centering
    \includegraphics[width=\linewidth]{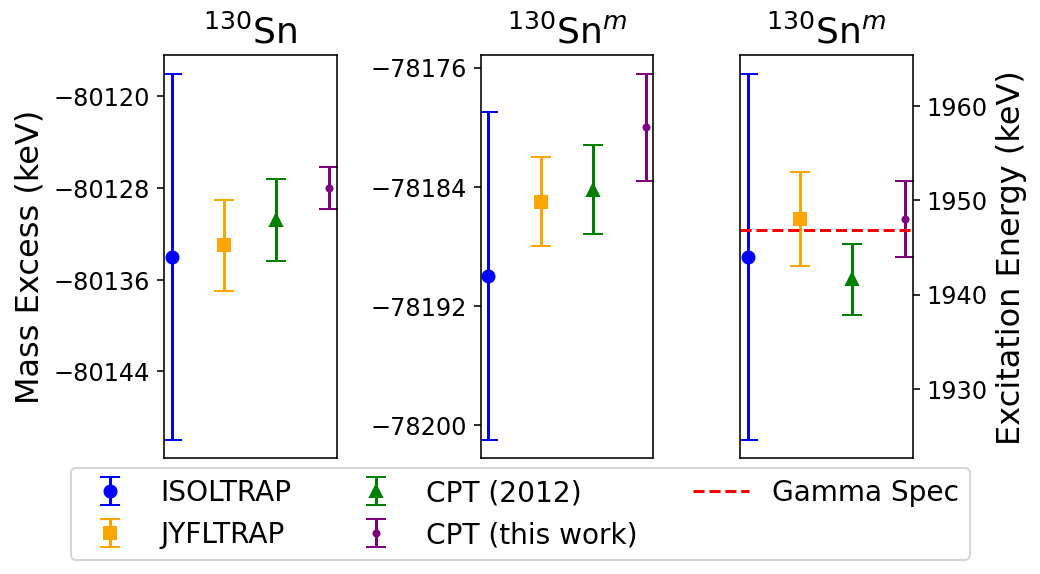}
    \caption{Comparison of recent CPT measurements (right most purple dots) with previous measurements of the same isotopes \cite{Redshaw2009,Scielzo2009,Rahaman2011,Nesterenko2012,Sikler2005,Hakala2012,Kankainen2012,VanSchelt2013,Schelt2012} The left and center panels give mass excesses while the right panel the excitation energy of the isomeric state. $^{130}Sn^m$ excitation energy from gamma spectroscopy is indicated by a dashed line \cite{1981530}}
    \label{fig: prev meas}
\end{figure}

The results of our measurements are shown in Table \ref{tab: masses}. The results for the \textsuperscript{130}Te mass excess were consistent with the more precise AME2020 value that comes from a highly precise measurement using the FSU trap \cite{Redshaw2009}. It also agrees with the $\textsuperscript{130}\text{Te}\rightarrow\textsuperscript{130}\text{Xe}$ Q-value from the CPT \cite{Scielzo2009}, JYFLTRAP \cite{Rahaman2011}, and SHIPTRAP\cite{Nesterenko2012}. Figure \ref{fig: prev meas} indicates that the \textsuperscript{130}Sn mass excess is consistent with all previous Penning trap measurements from ISOLTRAP \cite{Sikler2005}, JYFLTRAP \cite{Hakala2012, Kankainen2012}, and the CPT\cite{VanSchelt2013,Schelt2012} while being 2 times more precise. Similarly, the mass excess of the isomeric state is also consistent with past measurements. Finally, the excitation energy is in excellent agreement with the results from ISOLTRAP and JYFLTRAP while also agreeing with the very precise value from gamma spectroscopy \cite{1981530}.

\section{\label{sec:level1}R-Process Simulation}

To evaluate how the mass measurements from this work affect final r-process abundances, we first determined the best conditions to reproduce the solar r-process abundances by scanning the same parameter space as that explored in Ref. \cite{Meyer1997} within SkyNet \cite{Lippuner2017}. Namely, with an initial temperature of $9$ GK, we scanned initial electron fractions, $Y_e$, between $0.1$ and $0.5$, where $Y_e$ is defined as:
\begin{equation}
    Y_e=\frac{n_p}{n_p+n_n},
\end{equation}
where $n_p$ is the proton number density, and $n_n$ is the neutron number density. Since we assume the stellar environments are electrically neutral, $Y_e$ is a measure of neutron density. A small $Y_e$ therefore means that there are much fewer protons than neutrons in the environment, leading to heavier element production. We scanned initial entropies per baryon between $50$ $k_B$ and $500$ $k_B$. The entropy per baryon represents how homogeneous the system is with higher entropies meaning the environment is more homogeneous. In general, smaller entropies lead to more heavy element production. Finally, we scanned expansion timescales, $\tau$, between 1 ms and 1 s. The expansion timescale is the parameter which characterizes the rate of expansion according to a density evolution of $\rho\propto e^{-t/\tau}$. A higher timescale means a slower expansion. Therefore, there is a less violent explosion, allowing neutron capture seed nuclei to remain in the high neutron flux environment longer, making heavier elements.

However, unlike the calculations in \cite{Meyer1997}, SkyNet permits the entropy to change throughout the expansion, allowing for changes in entropy in accordance with the second law of thermodynamics.
To determine the set of initial conditions that would best reproduce the solar r-process abundances, we calculated the reduced $\chi^2$ of each final abundance pattern. To find the uncertainty for each initial condition, we note that, for a three parameter fit, a difference of 3.53 of the minimized reduced $\chi^2$ in any direction within the parameter space corresponds to a $1\sigma$ uncertainty \cite{Workman2022}.

\begin{figure}[h]
    \centering
    \includegraphics[width=\linewidth]{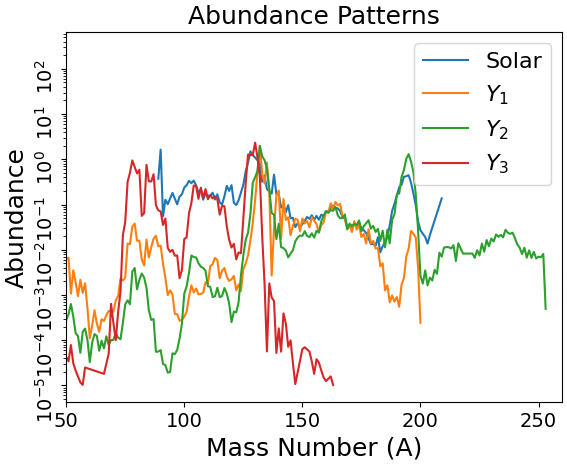}
    \caption{Final solar r-process abundance patterns of SkyNet calculations. These are compared to solar abundance data shown by the blue line. The fit to $Y_3$ ($Z<56$) is given by red line. The $Y_2$ ($Z>75$) fit is given by the green line. The fit to the total abundance pattern, $Y_1$, is shown by the orange line. The model fit parameters are summarized in Table \ref{tab: cond}}
    \label{fig: peak fits}
\end{figure}

After initial scans of the parameter space, the conditions that most closely reproduce the solar r-process abundances were a $Y_e$ of $0.3725(25)$, an expansion timescale of $210(5)$ms, and an entropy per baryon of $435(5)k_B$. We will call the abundance pattern resulting from these conditions $Y_1$. The final reduced $\chi^2$ for these conditions was 135, far from a good fit. As shown in Figure \ref{fig: peak fits}, these conditions reproduce the rare-earth section of the abundance pattern near $A=165$ well, but the fit diverges from the solar abundance pattern for both light and heavy mass numbers. 

To diminish the reduced $\chi^2$ and create a better fit, we noted that the abundance pattern can be divided into three sections: the $N=82$ peak ($A<135$), the rare-earth peak ($135\leq A\leq 186$), and the $N=126$ peak ($A>186$). $Y_1$ corresponds to the universality of the abundance of the observed r-process of elements in the range of $56<Z<75$ described observed in metal-poor stars of the Galactic halo \cite{Otsuki2003}. However, many models can predict the same abundances in this region but diverge outside of it. The results from fitting to this region of universality supports the conclusion in \cite{Otsuki2003,Shibagaki_2016,Yamazaki_2022} that r-process elements are not all produced in a single unique environment. The lack of any good fit to the whole abundance pattern in our parameter space further supports this conclusion.

\begin{table*}[t]
    \caption{Summary of peak fit conditions. $Y_1$ has high temperature and high entropy which fits the entire solar r-process abundance pattern the best. $Y_2$ has high temperature and low entropy which best reproduces the heavy peak. $Y_3$ has low temperature and low entropy which best reproduces the light elements in the solar r-process abundance pattern}
    \begin{ruledtabular}
    \begin{tabular}{ccccccc} 
    Fit & Temperature (GK) & $Y_e$ & Entropy ($\frac{k_B}{\text{baryon}}$) & Expansion Timescale (ms) & Relative Frequency & Impact Parameter\\ 
    \hline
    $Y_1$ & 9 & 0.3725(25) & 435(5) & 210(5) & 0.408(5) & 0.026\\  
    $Y_2$ & 9 & 0.2225(150) & $42.5_{-7.5}^{+12.5}$ & 7(3) & 0.168(4) & 0.010\\
    $Y_3$ & 5 & $0.26(15)$ & $55_{-30}^{+60}$ & $200_{-30}^{+50}$ & 0.424(5) & 0.222\\
    \end{tabular}
    \end{ruledtabular}
    \label{tab: cond}
\end{table*}

In an attempt to produce a good fit to the whole solar abundance pattern, we fitted each pattern from our parameter space to the regions of $Z<56$ and $Z>75$ individually to find the conditions that best reproduce these peaks. These results are shown in Figure \ref{fig: peak fits}.

While there were no good fits to the $Z<56$ region in our original parameter space, the conditions that best reproduced the region of $Z>75$, henceforth $Y_2$, were $\tau=7(3)$ms, $Y_e=0.2225(150)$, and an entropy per baryon of $42.5_{-7.5}^{+12.5}k_B$. While outside of our original parameter space, we found this minimum via a reduced $\chi^2$ gradient ascent algorithm. 

Our parameter space, being initially at 9 GK, consistently underproduced the first region. This suggests that this region is best explained by a colder low-entropy r-process which produces lighter nuclei. Therefore, we looked for light r-process conditions for this region. The light r-process is thought to occur at around 5 GK \cite{Psaltis2022}, and ideal conditions for this have already been identified \cite{Surman2014}. Considering the three cases presented in \cite{Surman2014} and performing a gradient ascent to find the minimum, we identified the parameters, henceforth $Y_3$, of $\tau=200^{+50}_{-30}$ms, $Y_e=0.25(15)$, and entropy per baryon of $55^{+60}_{-30}k_B$.

\section{Impact of New Masses on the R-Process}

To quantify the effect of our mass measurements, we use the impact parameter,
\begin{equation}
    \label{eq: impact parameter}
    F=100\sum_A|X(A)-X_b(A)|,
\end{equation}
where $X(A)$ are the original mass number abundances and $X_b(A)$ are the mass number abundances with the new mass measurements. The updated mass inputs had an impact for these conditions of 0.021 and led to a nearly identical reduced $\chi^2$. The negligible change in the reduced $\chi^2$ and small impact parameter is to be expected due to the small mass changes for species not directly on the r-process path. 

\begin{figure}[h]
    \centering
    \includegraphics[width=\linewidth]{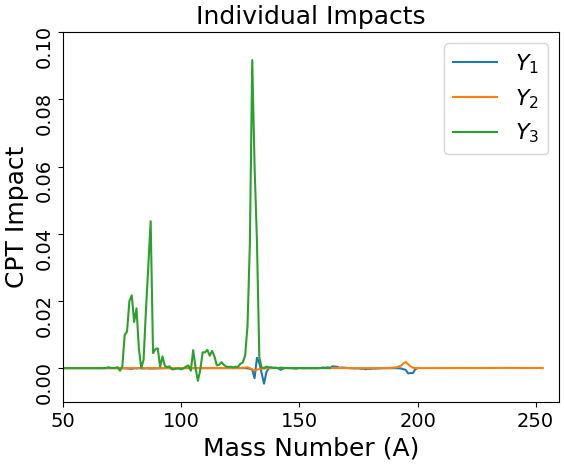}
    \caption{Individual changes which contribute to the summed impact parameter, Equation \ref{eq: impact parameter}. The most significant impacts happen in $Y_3$ and $Y_1$ abundance pattern near $A=130$ while the only significant changes for $Y_2$ occur near A=200.}
    \label{fig: CPT}
\end{figure}

Further, investigating the impact of the new mass measurements for each of our peak conditions, we found that the updated \textsuperscript{130}Te and \textsuperscript{130}Sn masses had the largest impact on the light r-process, $Y_3$. The individual contributions to the summed impact parameter are shown in Figure \ref{fig: CPT}. From Eq \ref{eq: impact parameter}, positive values represent decreases in abundance, and negative values represent increases in abundance. The largest changes for both $Y_1$ and $Y_3$ were not surprisingly around A=130. There were only very small changes in $Y_2$, leading to the lowest impact parameter for these conditions, but interestingly. The larger observed effect on $Y_3$ signifies that the new masses mainly affect the cold r-process abundance scenario.

To further increase the quality of the fit to the solar abundance data, we made a linear combination of the abundance patterns resulting from our optimal conditions. The final fit was $0.4083(47)Y_1+0.1676(39)Y_2+0.4241(48)Y_3$. These coefficients can be thought of as the relative frequency of events with the associated conditions. The difference in the relative frequencies of each set of conditions can be explained by the differences in the mechanisms of heavy-element nucleosynthesis. $Y_3$, the coldest r-process, mainly produces the lighter r-process isotopes and has the highest relative frequency. $Y_2$ produces the heaviest r-process isotopes and has the lowest relative frequency. Therefore, individual $Y_2$ events happen less frequently and produce more heavy elements per event than $Y_1$ and $Y_3$.

\begin{figure}[h]
    \centering
    \includegraphics[width=\linewidth]{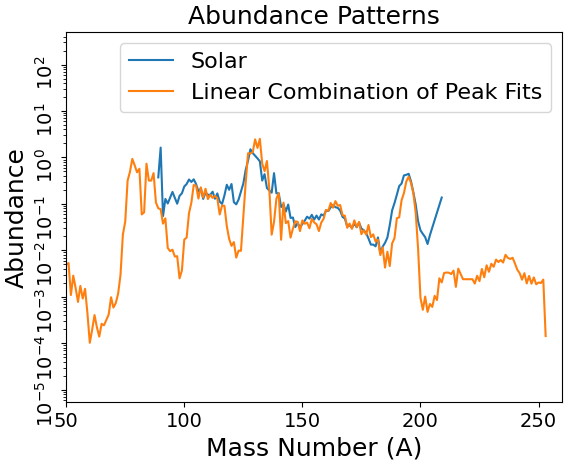}
    \caption{Final abundance pattern of the linear combination of SkyNet calculations with the conditions from Figure \ref{fig: peak fits} compared to solar r-process abundance data \cite{Arlandini1999}}
    \label{fig: combined fits}
\end{figure}

A summary of the conditions is shown in Table \ref{tab: cond}, and the final abundance of the linear combination fit is shown in Figure \ref{fig: combined fits}. This fit achieves an improved reduced $\chi^2$ of 92. 
\section{\label{sec:level1}Conclusion}
The mass excess of \textsuperscript{130}Te, \textsuperscript{130}Sn, and \textsuperscript{130}Sn\textsuperscript{m} have been measured for the first time using the PI-ICR technique with the CPT. The respective mass excesses of $-87354.4(21)$ keV, $-80128.0(18)$ keV, and $-78180.0(36)$ keV all agree with previous Penning trap measurements from ISOLTRAP \cite{Sikler2005}, SHIPTRAP \cite{Nesterenko2012}, CPT \cite{Scielzo2009,VanSchelt2013,Schelt2012} and JYFLTRAP \cite{Hakala2012, Kankainen2012, Rahaman2011} that used the TOF-ICR technique as well as with the FSU TRAP \cite{Redshaw2009} measurement that used the FT-ICR technique. The new \textsuperscript{130}Sn mass excess is also a factor of two more precise than the previously most precise value from JYFLTRAP \cite{Hakala2012}. The extracted excitation energy of 1947.5(34) keV for the measured \textsuperscript{130}Sn\textsuperscript{m} isomer is also consistent with the value obtained from gamma spectroscopy \cite{1981530}.

The new mass excesses were added to a SkyNet network calculation to assess their impact on three different r-process scenarios. These scenarios were chosen to best reproduce the entirety of the solar abundance ($Y_1$), heavy nuclei ($Z>75$) ($Y_2$), or light nuclei ($Z<56$) ($Y_3$). While scenarios $Y_1$ and $Y_2$ both occur at high temperature (9 GK), the latter presents a much lower entropy. Scenario $Y_3$ is both at low temperature (5 GK) and low entropy. The new mass
excesses were found to mainly affect scenario $Y_3$, which corresponds to a low entropy cold r-process that produces primarily light r-process nuclei.

So far, no single astrophysical scenario (and nuclear physics set of inputs) has managed to perfectly reproduce the solar r-process abundance pattern. The relative contributions from the scenarios $Y_{1-3}$, which cover
both cold/hot and low/high entropies, to the solar r-process abundance pattern were assessed by a $\chi^2$ minimization process. The main contributors were found to be the high-temperature, high-entropy and the low-temperature, low-entropy scenarios. Furthermore, a better representation of the solar abundance
data was obtained from the linear combination of these three scenarios as given by the 30\% reduction in the reduced $\chi^2$ of the fit to solar data.

\rule{\linewidth}{0.5pt}
\section{\label{sec:level1}Acknowledgments}
This work was performed with the support of US Department
of Energy, Office of Nuclear Physics under Contract No. DE-AC02-
06CH11357 (ANL), Nuclear Theory Grant No. DE-FG02-95-ER40934, the Natural Sciences and Engineering Research
Council of Canada, Canada under Grant No. SAPPJ-2018-00028, and
the US National Science Foundation under Grant No. PHY-2310059.
This research used resources of ANL’s ATLAS facility, which is a DOE Office of Science User Facility.

\end{document}